\documentstyle[12pt]{article}
\def\hybrid{\topmargin -20pt    \oddsidemargin 0pt
        \headheight 0pt \headsep 0pt
        \textwidth 6.25in       
        \textheight 9.5in       
        \marginparwidth .875in
        \parskip 5pt plus 1pt   \jot = 1.5ex}

\hybrid
\newskip\humongous \humongous=0pt plus 1000pt minus 1000pt
\def\caja{\mathsurround=0pt}
\def\eqalign#1{\,\vcenter{\openup1\jot \caja
        \ialign{\strut \hfil$\displaystyle{##}$&$
        \displaystyle{{}##}$\hfil\crcr#1\crcr}}\,}
\newif\ifdtup

\relax

\def\be{\begin{equation}}
\def\ee{\end{equation}}
\def\ba{\begin{eqnarray}}
\def\ea{\end{eqnarray}}

\begin{document}
\renewcommand{\theequation}{\thesection.\arabic{equation}}
\newcommand{\beq}{\begin{equation}}
\newcommand{\eeq}[1]{\label{#1}\end{equation}}
\newcommand{\ber}{\begin{eqnarray}}
\newcommand{\eer}[1]{\label{#1}\end{eqnarray}}
\begin{titlepage}
\begin{center}

\hfill CPTh-S369.0895 \\
\hfill Crete-95-16 \\
\hfill \\

\vskip .5in

{\large \bf  MEMBRANES
IN THE TWO-HIGGS STANDARD MODEL }
\vskip .5in

{\bf C. Bachas} \footnotemark \\

\footnotetext{e-mail address: bachas@orphee.polytechnique.fr}

\vskip .1in

{\em Centre de Physique Th\'eorique\\
Ecole Polytechnique \\
 91128 Palaiseau, FRANCE}

\vskip .15in

       and

\vskip .15in

{\bf T.N. Tomaras} \footnote{e-mail address:
tomaras@plato.iesl.forth.gr }\\
\vskip
 .1in

{\em  Physics Department, University of Crete\\
and Research Center of Crete\\
 714 09 Heraklion, GREECE
  }\\

\vskip .1in

\end{center}

\vskip .4in

\begin{center} {\bf ABSTRACT }
\end{center}
\begin{quotation}\noindent
We present some non-topological
static  wall solutions  in two-Higgs
extensions of the standard model. They
are classically-stable
in a large region of parameter space,  compatible with
perturbative unitarity and with present phenomenological bounds.

 \end{quotation}
\vskip1.0cm
August 1995\\
\end{titlepage}
\vfill
\eject

\setcounter{equation}{0}

There are several reasons to consider extensions of the Higgs
sector of the standard model, most notably
the introduction of an extra source of soft
 CP violation \cite{vivlio},
the possibility of generating sufficient baryon asymmetry at the
electroweak transition \cite{Baryon},
and low-energy supersymmetry.
An extended Higgs sector allows for the possibility
of discrete symmetries and of associated
domain walls \cite{Zeld,Vil}.
In this letter we will present another class of
membrane defects in the two-Higgs standard model
that differ from domain walls in two important
ways: ({\it a})  they are not tied to a discrete
symmetry  and are thus more generic, i.e. they
  exist in a codimension-zero region of parameter space, and
({\it b}) they are classically- but not topologically-stable
and have a finite, though possibly cosmologically-long
life time. They resemble in these respects
the previously discussed $Z$ strings \cite{Z},
while contrary to these latter \cite{Z1}
they are as we will show stable in a realistic range  of parameters.
This range does not however include the minimal supersymmetric
standard model.
The new defects are embeddings of some  recently found \cite{Mex}
solutions of the
$2d$ Abelian-Higgs model with two or more complex scalars
\footnote{
The 2d Abelian-Higgs model has
a plethora of other sphaleron-like solutions
\cite{sphal} which can be embedded similarly in the two-Higgs
standard model. Since these are classically unstable they will
be of no concern to us in this letter.}
{}.
They are characterized by the non-trivial winding of a relative
U(1) phase of two Higgses in the direction ($x$) normal to the
wall and are electrically neutral.
Their energy per unit area
and thickness  are of order $m_W^2 m_A/\alpha$ and
$m_A^{-1}$ respectively, where $\alpha$ is the fine-structure constant
and $m_A$ the mass of the CP-odd Higgs scalar.
\vskip 0.2cm

The nature of these defects is best illustrated by a complex-scalar
field theory in four dimensions with potential
$ V(\Phi) = {\lambda\over 4} (\Phi^* \Phi - v^2)^2 - \mu^2 v Re\Phi$.
This has a unique minimum at a real value of $\Phi$, so that
  there are no
 topologically-stable domain walls.
Nevertheless, it can be shown \cite{Mex,unpub}  that
 for $\sqrt{2\lambda} v/\mu \geq 6.1 $
 there exists a classically-stable static wall solution characterized by
the
fact that the phase of $\Phi$ changes
by $2\pi$ as $x$ varies from $-\infty$ to $\infty$.
In the $\lambda\rightarrow\infty$ limit the  solution reduces to
 the well-known 2d sine-Gordon soliton,
 while for generic values of
$\lambda$ it can be analyzed numerically or via a $1/\lambda$
expansion. One can also study some features of the wall
analytically \cite{Mex} by trading the $\mu$ term in the potential with
periodic conditions in the $x$ direction. Notice indeed that the
  $\mu$ term   lifts the U(1) vacuum degeneracy   thereby forcing
the field to come back to its minimum within a distance
(wall thickness)  $\Delta x
\sim \mu^{-1}$. Alternatively we can achieve
the same result by making space
into a cylinder of radius  $L\sim \mu^{-1}$.
We will use this technical stratagem in the sequel, but we should
stress that the existence and stability of the membranes is {\it not}
tied to the existence of any accidental global symmetry.
\vskip 0.2cm

The Lagrangian of the two-Higgs standard model is
$$ {\cal L} =
  -{1\over 4} W^a_{\mu\nu} W^{a\mu\nu} -{1\over 4} Y_{\mu\nu}
Y^{\mu\nu}
+ \vert D_\mu H_1\vert^2 + \vert D_\mu H_2\vert^2
-V(H_1,H_2) \
\eqno(1)
$$
where
 $ W^a_{\mu\nu}=\partial_\mu W^a_\nu - \partial_\nu W^a_\mu - g \epsilon^
{abc} W^b_\mu W^c_\nu $ and
$ Y_{\mu\nu}=\partial_\mu Y_\nu - \partial_\nu Y_\mu
 $, the physical $Z^0$ and photon fields are
$Z_\mu =  W^3_\mu {\rm cos}\theta_W - Y_\mu {\rm  sin}\theta_W$ and
$A_\mu =  W^3_\mu {\rm sin}\theta_W + Y_\mu {\rm  cos}\theta_W$
and
${\rm tan} \theta_W = g^\prime/ g$.
Both Higgs doublets have hypercharge equal to one,
the covariant derivative is
$$
D_\mu H_I=(\partial_\mu + {i\over2} g \tau^a W^a_\mu + {i\over2} g^\prime
Y_\mu) H_I
$$
for $I=1,2$  and the  potential reads
 $$\eqalign{
 V(H_1,H_2)& = \lambda_1 \Bigl( \vert H_1\vert^2 - {v_1^2\over2}\Bigr)^2 +
 \lambda_2 \Bigl( \vert H_2\vert^2 - {v_2^2\over 2}\Bigr)^2 +
\lambda_3 \Bigl( \vert H_1\vert^2  + \vert H_2\vert^2 -
{v_1^2+v_2^2\over2}\Bigr)^2
\cr
&+ \lambda_4 \Bigl[ \vert H_1\vert^2 \vert H_2\vert^2 - (H_1^\dagger
H_2)  (H_2^\dagger H_1)\Bigr]
 + \lambda_5 \Bigl[{\rm Re }
(H_1^\dagger H_2) -{v_1 v_2\over2} cos\xi\Bigr] ^2 \cr
& \hspace{3cm} +
\lambda_6 \Bigl[ {\rm Im}( H_1^\dagger H_2)-{v_1 v_2\over2} sin\xi\Bigr]^2
\cr}
\eqno(2)
$$
where $\vert H_I \vert^2 \equiv  H_I^\dagger H_I$.
This is the most general potential \cite{vivlio}
 subject to the condition that
both CP invariance and a discrete $Z_2$ symmetry ($H_1\rightarrow -H_1$)
are only broken softly. The softly broken $Z_2$ symmetry is there
to suppress unacceptably large
flavor-changing neutral
currents.
Assuming all the $\lambda_i$ are
positive, the minimum of the
potential up to a gauge transformation is at
$$
<H_1> =  \  e^{-i\xi} \left(  \matrix{ 0 \cr  v_1/ \sqrt{2} \cr} \right)
\ \ {\rm and} \ \
<H_2> =  \left( \matrix{ 0 \cr { v_2/ \sqrt{2}}\cr } \right) \ .
\eqno(3)
$$
To reduce the large number of parameters we will restrict ourselves to
$\lambda_1=\lambda_2$, $\lambda_5=\lambda_6$, $\xi = 0$ and
$v_1=v_2=v$ (or ${\rm tan}\beta = 1$) in the sequel.
Relaxing these conditions is straightforward but beyond the scope
of the present letter. In addition to the electroweak gauge bosons
with masses $m_W^2= g^2 v^2/2$ and $m_Z = m_W/cos\theta_W$ ,
 the perturbative
spectrum contains
a charged Higgs boson $H^+$ with
mass $m_{H^+}^{\ 2} = \lambda_4 v^2 $ ,
a CP-odd neutral scalar $A^0$ with mass $m_{A}^{\ 2}= \lambda_5 v^2 $ ,
and two CP-even neutral scalars $h^0$ and $H^0$ with masses
$m_h^{\ 2} = 2\lambda_1 v^2$ and
$m_{H^0}^{\ 2} = (2\lambda_1+4\lambda_3+\lambda_5)  v^2$ respectively.
\vskip 0.2cm

For $\lambda_5=0$ there is an accidental U(1) global symmetry with
$A^0$  the associated Goldstone boson.
The corresponding phase is the phase of interest that
winds around non-trivially when   crossing
the membrane. Using the previously-mentioned stratagem we will first
work in this  $m^2_{A}=0$  limit but compactify
 space into a cylinder
of period $2\pi L$ in the $x$ direction.
We will also work in the temporal $Y_0 = W_0^a = 0$ gauge.
The following configuration is then the relevant   static,
 $y-$ and $z-$independent solution
of the classical   equations of motion
$$
H_1 =  e^{ix/L} \left( \matrix{ 0 \cr  F \cr } \right) \ ,
\ \
H_2 =  \left( \matrix{ 0 \cr  F\cr } \right) \  \ ,
\ \
Z_x = {cos\theta_W \over g L} \ \
{\rm and} \ \ A_x = {a\over L} \ \eqno(4)$$
  where
$$F^2 =  {v^2\over 2}\Bigl( 1 - {1\over 2{m_{H^0}^2}L^2 }\Bigr) \ ,
\eqno(5)
$$
$a$ is an arbitrary constant
\footnote{Note that $a$ is an angular variable since large gauge
transformations can change it by integers.
The possibility of this Wilson-line background is an
artifact of our stratagem and has
no analog in the non-compact case. Its net effect is to weaken the
stability
under  charged-field fluctuations as we will show shortly.},
 and   all other fields are equal to
zero.
The energy per unit
area (${\cal A}$)
  can be computed easily with the result
$$ E / {\cal A} = {m_W^2 sin^2\theta_W \over 4\alpha L}
\Bigl( 1 - {1\over 4{m_{H^0}^2}L^2} \Bigr) \ .  \eqno(6)
$$
It vanishes in the limit ($L\rightarrow\infty$)
of a very thick membrane.
Note   that since the charged upper components of the
Higgs fields as well as the charged gauge bosons
${W^{\pm}}_\mu = (W^1_\mu \mp i W^2_\mu)/\sqrt{2}$ vanish,
the membrane has no electromagnetic couplings.

To check for classical  stability
 we may restrict
ourselves to  $x$-dependent
fluctuations and   vector fields in only the $x$ direction.
These are the fluctuations of the theory in one spatial
dimension.
Stability under such perturbations is of course a necessary
requirement,  but it is also a sufficient one for the following
reason: first, because of $y$- and $z$-translational invariance we may
diagonalize the fluctuations by going to the $(k_y,k_z)$
Fourier space. Second we may break the energy density
of any static field configuration  as
$E = \int ({\cal E} + {\tilde {\cal E}})$ where
$$
{\cal E} =
\vert D_x H_1\vert^2 + \vert D_x H_2\vert^2
+V(H_1,H_2)
\eqno(7a)
$$
and
$$ {\tilde {\cal E}}=     {1\over 4} Y_{ij} Y_{ij}
+ {1\over 4} W_{ij}^a W_{ij}^a  +  \sum_{I=1,2}
 \Bigl( \vert D_y H_I\vert^2 + \vert D_z H_I\vert^2 \Bigr) \ . \eqno(7b)
$$
Now the quadratic fluctuations of ${\cal E}$ are
independent of $(k_y,k_z)$ and of the $y$- and $z$- components of
the vector
fields,
 while ${\tilde {\cal E}}$ is a positive
semi-definite contribution which vanishes when $k_y=k_z=0$ and when all
vectors point in the $x$ direction.
This proves that
 stability of any static 2d soliton   guarantees the
stability of the corresponding 4d wall solution.

Next we note that the fluctuations of electrically-charged
and neutral fields do not mix at the quadratic level, so we can
study these two sets of fields separately.
Using the residual invariance under $x$-dependent gauge transformations
we can go to a   gauge  in which the upper component of the
second Higgs doublet ($H^+_2$) is zero.
The fluctuations of the remaining charged fields can be
  Fourier
decomposed as follows:
$ g F W_x^+/ \sqrt{2} = - \sum \alpha_n e^{-inx/L}$ and
$ H_1^+ = e^{ix/L} \sum \beta_n  e^{-inx/L}$.
Using the form of the covariant derivative
$$
D_\mu H_I = \Biggl[ \partial_\mu + ig \left(
\matrix{ A_\mu sin\theta_W  + Z_\mu sin\theta_W cot({2\theta_W})
& {W_\mu^+}/\sqrt{2}  \cr
{W_\mu^-}/\sqrt{2}
& - {Z_\mu}/{2 cos\theta_W} \cr} \right)  \Biggr]
\ \left( \matrix{H^+_I\cr H^0_I \cr} \right) \eqno(8)
$$
one finds  after some straightforward algebra
the following variation of the energy per unit wall area
$$
\delta E /{\cal A} =  \ \sum_{n=-\infty}^{+\infty}
\left( \matrix{ \alpha_n^*  &\beta_n^* \cr} \right)
\left(
\matrix{ 2 & n-{\tilde a}  \cr
n-{\tilde a} & (n-{\tilde a})(n+1 -{\tilde a}) + \lambda_4 F^2
  \cr } \right)
\left( \matrix{\alpha_n  \cr {\beta_n}
 \cr } \right) + {\rm cubic}
\eqno(9)
$$
with ${\tilde a} = 1+ cos^2\theta_W + ag sin\theta_W $.
This is manifestly positive-definite for any ${\tilde a}$
if and only if
$$ 2\lambda_4 F^2 L^2  =   m_{H^+}^2 L^2 ( 1 - {1\over 2m_{H^0}^2 L^2})
> 1 \ . \eqno(10)$$
The strongest conditions are in fact obtained
for ${\tilde a}={\rm integer}$
  corresponding to a
vanishing $W_x^3$ background.
 Considering
next  the electrically-neutral sector we note first
that up to the constant
Wilson-line background $a$, the photon  field $A_x(x)$ is a pure gauge. The
fluctuations of the remaining fields $Z_x$, $H_1^0$ and $H_2^0$ are   those
analyzed in ref.\cite{Mex} in the context of the Abelian-Higgs
model with two complex scalars.  Stability under these fluctuations
yields one extra condition on the parameters of the model
$$
  4\lambda_1 F^2 L^2  =   m_{h}^2 L^2 ( 1 - {1\over 2m_{H^0}^2 L^2})
> 1 \ . \eqno(11)
$$
 Taken together inequalities (10) and (11) tell us that
for  the winding
solution to exist and be classically stable,
  all but the CP-odd scalar
must be sufficiently massive
in units of the cutoff or  ``membrane thickness"  $L$.
Notice that
since the problem is effectively two-dimensional the gauge couplings
do not enter into the stability conditions.

Now in the realistic case of a
non-compact space we expect $m_A^{-1}$   to replace
  $L$ in the above discussion.
This can be seen explicitly in
 the $\lambda_1, \lambda_3, \lambda_4 \rightarrow
\infty$ limit in which all but the CP-odd scalar $A^0$
are infinitely massive. Finiteness of the potential
energy constrains in this limit the
  Higgs doublets to   the form:
$$H_1 = e^{i\theta/2} \left(  \matrix{ 0 \cr  v/ \sqrt{2} \cr} \right)
\ \ {\rm and} \ \
H_2 = e^{-i\theta/2}  \left(  \matrix{ 0 \cr  v/ \sqrt{2} \cr} \right)\ ,
$$
modulo of course a gauge transformation. The energy of any static
  configuration
  reads in this case
$ E  = \int   [ {v^2\over 4} (\nabla \theta)^2 - {v^2\over 2}
m_A^2 cos\theta  ] + {\rm gauge}$
where the gauge contribution is minimum when $W_\mu^a = Y_\mu^a =0$.
 The above
energy functional admits the well-known sine-Gordon soliton solution
$$
\theta = 4 \ {\rm arctan}\Bigl( {\rm exp} (m_A x)\Bigr) \eqno(12)
$$
which describes in our context a topologically-stable wall
of thickness
$m_A^{-1}$ and   surface tension
$ E/{\cal A} = 4v^2m_A \equiv
2 m_W^2 m_A sin^2\theta_W/ \pi \alpha$.
 These results agree qualitatively with eqs. (4-6), if one replaces
$L$ by $m_A^{-1}$ and takes $m_{H^0}\rightarrow\infty$.

For finite   values of
$\lambda_1, \lambda_3$ and $\lambda_4$  we  have performed the
analysis numerically as follows:
 starting
 with an initial   configuration
close to the above sine-Gordon soliton
 we followed
 the direction of steepest descent until either
we arrived at the vacuum, or the change in energy  per step  was
less than one part in $10^{12}$. One step amounted typically to
a change of fields   $\sim 10^{-2}$ times the
gradient of energy. Stopping at a non-zero energy was interpreted
as evidence for  the existence of a stable solution.
We verified that these candidate solutions obeyed the virial relation
to better than one part in $10^3$, and that they were insensitive
to changes of the initial configuration or of the
cutoff on the convergence rate.
The entire numerical analysis was performed in the
 $W_x^a = Y_x =0$ gauge. The profile of a typical solution,
plotted in figures
  1 and 2 , differs little  from the
  sine-Gordon soliton. Likewise the
  energy per unit area   stayed  typically  within 10\% of
$4v^2 m_A$ .
 The conditions
for classical stability
were to within a few percent found to be
$$
m_h/m_A \geq 2.0 \;\;\;\; {\rm and} \;\;\;\;
 m_{H^+}/m_A \geq 2.2   \eqno(13)
$$
while the condition on $m_{H^0}/m_A$ was sensitive to
the precise value of the ratio $m_h/m_A$.
 The region of stability
for three selected values of this ratio is depicted in fig. 3.
As already anticipated  stable walls exist provided $A^0$ is
sufficiently light compared to all other scalars.
Taking $m_A \simeq 50 GeV$,   close to its experimental lower
bound, we may satisfy these constraints with
 scalar self-couplings   $\le 1/2$.
This is well within the
  region of perturbative unitarity in which the semi-classical
approximation can be trusted.
The minimal supersymmetric standard model,
on the other hand, lies
outside this region of stability,
as can be seen for instance
from the fact that   $A^0$ is not
the lightest neutral scalar\cite{vivlio}.
It is however conceivable that
 loop corrections  modify this conclusion.

The defects described in this letter would not interact
 electromagnetically, unless they happen to acquire charge by trapping
fermions. Due, on the other hand,  to their large
energy density
($E/{\cal A} \sim  10^{10}gr/cm^2$
assuming  $m_A\sim m_W$)
they would   manifest themselves through gravitational attraction.
 A single wall crossing today the entire
universe would for example overclose it
and can  be excluded.
Smaller membranes which either collapsed or were torn apart
by quantum tunneling
may   have acted as seeds for the formation of
galaxies. The mass of a typical galaxy is in fact comparable to that
of a membrane a few light years in size.

\vskip 0.3cm
{\bf Aknowledgements}
C.B. thanks the Physics Department of the
University of Crete and
the Research Center of Crete, and T.N.T.
thanks the Centre de Physique
Th\'eorique of the Ecole Polytechnique
for hospitality while this work was being
completed. This research was supported in part by the EEC grants
CHRX-CT94-0621 and   CHRX-CT93-0340, as well as by
the Greek General Secretariat
of Research and Technology grant $91E\Delta 358$.


\newpage

\vspace{2cm}

{\bf FIGURE CAPTIONS}

Figure 1: The profile of a typical membrane, for $m_h = 2.5$,
$m_{H^0} = 5.0$ and $m_{H^+}= 4.0$ in units of $m_A$. Both the
gauge fields and the charged upper components of the doublets vanish.
Plotted are the real and imaginary parts as well as the
magnitude squared of the   neutral components
$H^0_1$ and $H^0_2$, as functions of the coordinate $x$ normal to the
membrane.

 Figure 2: Plot of the neutral components of the two Higgs doublets
 in the complex plane.
Their phases wind around  in opposite
directions as one crosses the wall, starting at zero and joining
at $\pm\pi$.

Figure 3: The boundaries of classical stability in the
$(m_{H^0}, m_{H^+})$ plane for three different values of $m_h$.
The scale is chosen so that $m_A =1$.
Classically-stable membranes exist above the indicated lines.
Also given is the energy density in units of $v^2 m_A$,  for some
selected points close to the boundary.

\end{document}